\documentstyle{lamuphys}
\makeatletter
\let\chapter\hid@chapter
\makeatother
\begin{document}
\pagenumbering{arabic}
\title{Evolution of Normal Galaxies: HST Morphologies and Deep Spectroscopy}

\author{Richard Ellis}

\institute{Institute of Astronomy, Madingley Road,
Cambridge, UK}

\maketitle

\begin{abstract}

I review progress in understanding the evolution of normal field and
cluster galaxies through the combination of HST imaging and
ground-based spectroscopy. These data suggest that the bulk of the star
formation producing the present-day galaxy population occurred at
accessible redshifts, $z<$2. Furthermore, a surprising amount of the
detailed processing that shaped the Hubble sequence and
morphology-density relation occurred surprisingly recently. The stage
is thus set for a concerted attack on these questions with the present
generation of 8-10 metre large telescopes. An important step forward
will be the development of efficient survey techniques for the
systematic exploration of the $z>$1 Universe. Some possible approaches
are briefly discussed.
 
\end{abstract}
\section{Introduction and the Role of Instrumentation}

These are exciting times in the fields of observational cosmology,
galaxy formation and evolution. The rate of publications on bulletin
boards like SISSA is one (albeit possibly unreliable) indicator of
interest. At the time of writing almost a third of all 1996 preprints
thus far are in some way connected with these topics. One is also
struck by the optimism in many minds that we are close to resolving one
of the outstanding questions of modern cosmology, namely `When did
galaxies form?'. Lest we be over-optimistic about the progress we have
apparently made, some of which I review below, it is salutory to recall
Zel'dovich's enthusiastic remarks when he summarised the 1977 IAU
Symposium on `Large Scale Structure' in Tallinn. He claimed
`..extrapolating to the next symposium in the early eighties, one can
be pretty sure that the question of the formation of galaxies and
clusters will be solved'!

Although it pays to be cautious in reviews, with results in abundance
from the refurbished HST (not least from the remarkable Hubble Deep
Field), and the first deep spectroscopic studies emerging from the Keck
telescope, it is a timely moment to look ahead and discuss, in the
context of the VLT, the respective roles of HST and large ground-based
telescopes.

In planning programmes with the VLT, one must surely now take into
account the quite remarkable capabilities of HST. HST already offers
$\simeq$1 kpc image resolution at all redshifts in most conventional
cosmologies and to this will soon be added long-slit spectroscopy and
infrared imaging provided STIS and NICMOS are delivered according to
schedule.

HST is {\it not}, as one often sees quoted, a `small aperture
telescope'. At wavelengths where the ground-based sky is dominated by OH
emission, HST offers a significant gain. For background-limited work,
HST has an aperture equivalent to a 6.5m ground-based instrument.
Furthermore, the wavelength range where this advantage accrues
(0.8$<\lambda<$ 1.6 $\mu$m) is particularly crucial in studies of the
$z>$1 Universe. Likewise, although HST does suffer from having a small
field of view, it is not yet clear whether large aperture telescopes 
using adaptive optics will recover resolution approaching that of HST over 
a big enough field to make an appreciable gain. And in the longer term, 
HST's Advanced Camera will offer an important step forward.

Therefore, I would contend that the traditional `complementarity'
between HST and ground-based telescopes has been overstated and that
ground-based telescopes should, at least in this scientific area,
respond with imaginative instrumentation. Their big advantage over HST
is the ability, at least in principle, to respond rapidly to both
technological progress and scientific developments.  Rapid interplay is
important in both directions. One example of science driving
instrumentation might be the discovery that the remarkably small angular
sizes of the majority of star forming sources in the Hubble Deep Field has
significant implications for ground-based instruments designed to study
them. An example of technology driving scientific capability is the
recent availability of large format HgCdTe arrays which opens up the
prospect of survey spectroscopy at high redshifts.
Such interaction should ensure ground-based telescopes maintain a
cutting edge in years to come.

Yet, with a few exceptions, much of the instrumentation being developed
for the 8 metre telescopes coming online is of the `monumental'
variety. By this I mean large \$3-5M general purpose instruments which
have taken many years to develop and, because of the cost and
associated wide community involvement, often become the brainchilds of
committees preoccupied primarily with technical reliability and
financial management. Although important considerations, I'm confident
many would agree that we should ensure at least some funds are set
aside for instrumentation that can be developed quickly to exploit what
is, after all, a rapidly moving subject.

This article, therefore, addresses the progress made in understanding
the evolution of normal galaxies from both HST and ground-based
telescopes. \S2 discusses the progress made in the study of cluster spheroidal
galaxies together with the important implications these results may have
on the visibility of distant primordial sources. I also make some remarks on
the environmental effects occurring in clusters. \S3 discusses results
on the evolution of field galaxies, particularly in the context of the
faint blue galaxy problem and deep imaging from HST. Finally, in \S4 I
speculate briefly on ways in which we might systematically explore the 
$z>$1 Universe.

\section{The Star Formation History of Cluster Galaxies}

Ellipticals were traditionally imagined to be simple stellar systems 
whose stars were formed in a single burst of star formation at high redshift 
(Tinsley \& Gunn 1976, Sandage \& Visvanathan 1978, Bower et al 1992).
The popularity of this hypothesis is easy to understand. Such systems
are convenient for theorists to model and, as their ancestors should
be luminous primaeval galaxies, the hypothesis produce exciting 
opportunities for observers too!

However, in the past decade, the simple picture has been under concerted
attack. Many local ellipticals show evidence for intermediate-age 
stellar populations (O'Connell 1980) and dynamical peculiarities 
seen in many (dust-lanes, shells etc) can be readily explained if
they formed more recently via the merger of gas-rich systems (Toomre 1977, 
Quinn 1984). Such arguments suggest a continued formation of ellipticals
to quite low redshifts.

The conflict might be resolved if some ellipticals were old single
burst systems, whereas the remainder formed via merging of gas-rich disk
galaxies. In this case one might expect an environmental and/or mass
dependence in the rate of occurrence of intermediate age populations.
Reasonably good evidence is emerging that recent star formation is more
prevalent in low density environments than in clusters. Rose et al
(1994) find the mean stellar dwarf/giant ratio is higher in environments
with low virial temperatures. This would be consistent with other
environmental trends which indicate accelerated star formation histories
in clusters (Oemler 1991). Kauffmann et al (1996) have suggested deep
field redshift surveys indicate a paucity of high $z$ red spheroidals,
although the reliability with which such systems can be identified using
ground-based colours needs to be verified with HST data.
  
The sensitivity of the $U$-band light to small numbers of hot, young stars
enabled Bower et al (1992) to conclude that no more than 10\% of the
current stellar population in present-day E/S0s could have been formed
in any subsequent activity in the past 5 Gyr as might be the case if
merging of gas-rich systems had been involved. This result presents an
important challenge for hierarchical theories of structure based on dark
matter halos since these predict relatively recent formation eras for
massive galaxies. Kauffman (1996) and Baugh et al (1996) have addressed
the question quantitatively using a simple prescription for
merger-induced star formation. They find that the homogeneity of Bower et al's
colour-magnitude (c-m) data can be satisfied if the merging of disk 
galaxies that produce spheroidals was largely complete by a redshift
$z\simeq$0.5.

Although good progress has been made in tracking the UV-optical
c-m relation to higher redshift (Ellis et al 1985, Arag\'on-Salamanca et
al 1991, 1993), without morphological information a major uncertainty
remains. The scatter of the photometric c-m relation may be
underestimated if some spheroidal galaxies lie blueward of the c-m
sequence. This could well be the case if the timescale for dynamical
evolution is shorter than that for main sequence evolution as indicated
in numerical simulations (Mihos 1995, Barger et al 1996a). 

The MORPHS team (Dressler et al 1996) have recently extended the
analysis of Bower et al to a sample of three $z\simeq$0.54 clusters,
taking advantage of HST to morphologically classify a sample of 177
faint spheroidal galaxies (Ellis et al 1996a). The clusters cover a range
of optical richnesses and X-ray luminosities within a narrow redshift
interval specifically chosen so that observed colours are close to
rest-frame $U-V$. Overall, the morphological selection of Es appears
reliable to $I$=23. However, the distinction between E and S0 galaxies
becomes somewhat uncertain fainter than $I$=21-22.

\begin{figure}
\vspace{6.5cm}
\caption{Colour-magnitude diagram for morphologically-classified
galaxies in 0016+16 ($z$=0.54) from the `MORPHS' project (Ellis et al 1996a).
Es are indicated by filled circles, S0's by triangles and E/S0s by squares.
Those spheroidals and compact objects known to be field galaxies or 
discounted from the analysis are indicated by open circles.The small scatter 
in the rest-frame $U-V$ colours of the spheroidal population in this and 
other clusters argues that the bulk of the stars formed before a redshift 3.} 
\end{figure}

The rest-frame $U-V$ colour-magnitude relations (Figure~1) for the
morphologically-confirmed spheroidals in these clusters show remarkably
small scatter ($<$0.1 mag rms) and there is no evidence the scatter for
S0s is any larger than that for Es. After accounting for photometric
errors, the intrinsic scatter is about 0.07 mag uniformly to
$I$=23 ($M_V=-$17.8 + 5 log$h$).  Moreover, the combined sample
shows little evidence of cross-cluster differences at a level greater
than the internal scatter. The most straightforward interpretation is
that the bulk of the star formation in cluster spheroidals occurred at
least 5 Gyr before a redshift of $z$=0.54, i.e. $z>$3 unless $H_o$ is
low or $\Lambda\neq$0.

Although this result is consistent with analyses of larger samples of
distant clusters (Arag\'on-Salamanca et al 1993), it does not
necessarily apply generally to {\it all} elliptical galaxies, even those
in clusters. Franx \& van Dokkum (1996) warn of selection effects
that might operate if ellipticals are identified morphologically in ways
that guarantee they are least 2-3 Gyr old at any redshift. The most
robust statement that can be made is that the stars that form the
dominant proportion of red light in 3 $z$=0.54 clusters most likely
formed before $z\simeq$3. Thus there is every incentive to search for
the star-forming ancestors of these galaxies.

Some of the above caveats might be minimised by examining the evolution
of spheroidal galaxies in terms of mass/light ratios rather than
broad-band luminosities. Impressive progress has been made of late in
measuring stellar velocity dispersions (Franx \& van Dokkum 1996, Bender
et al 1996) and HST scale sizes (Pahre et al 1996, Barger et al 1996b)
for high redshift galaxies. Preliminary results indicate only modest
evolutionary changes consistent with passive evolution from a burst of
star formation at high $z$. However, the selection biases above will
only ultimately be overcome with a comprehensive sample of field
spheroidals studied in a variety of ways. That Kauffmann et al (1996)
and Lilly et al (1995) should come to rather different conclusions from
analyses of the same CFRS dataset on the rate of evolution of field
ellipticals is an indication of the degree of uncertainty inherent in
the presently-available small samples.

How does the above help us to understand the physical origin of the
morphology-density relation (Dressler 1980) which, according to
observational evidence, was produced at quite low redshifts by
environmental effects (Butcher \& Oemler 1978, Allington-Smith et al
1993)? Morphological surveys of distant clusters such as those discussed
above (Couch et al 1994, Dressler et al 1994) delineate a clear change
in the morphological mixture in the sense that the proportion of disk
galaxies was much greater in the past, apparently at the expense of a
declining S0 population. 

One traditional explanation for this evolution, viz. the transformation
of spirals to S0s, goes a long way towards explaining the HST results
(Dressler et al 1996). The cluster ellipticals provide a backbone of
stability over a large range in redshift. By contrast, gas-rich spirals
enter the cluster potential and are stripped to produce the abundant
S0s we see in present day clusters. The evidence of radial gradients in
diagnostic spectral features is particularly convincing support of this
picture (Abraham et al 1996a). On the other hand, the small scatter
seen in the S0 population at all epochs thus far studied is puzzling
and, at least in the core regions, there appear to be some genuine
ellipticals which have surprisingly strong $H\delta$ absorption lines
indicating recent star formation (Barger et al 1996a).

A worry with all these studies thus far is the absence of a clear
understanding of how the clusters were selected.  Kauffmann (1996)
argues that, by selecting the richest clusters at a given redshift, we
are unlikely to be studying the precursors of present-day clusters. An
X-ray flux-limited sample may not be much better given our limited
physical understanding of the evolution of the X-ray luminosity
(Castander et al 1995). Ultimately, one might contemplate undertaking a
comprehensive survey using gravitational lensing to locate mass in a
well-defined manner. At that stage of complexity, it is probably
simpler to undertake very large field surveys if the primary goal is to
understand the galaxy population.

\section{Evolution of Field Galaxies}

The surveys of Bergeron \& Boiss\'e (1991) and Steidel et al (1994) based
on the identification of the galaxies responsible for Mg II absorption
in QSO spectra indicate little change in the overall luminosity function 
(LF) of regular field galaxies to $z\simeq$0.7. However, several details
remain unclear with the interpretation of such samples in the context
of galaxy evolution. These include the weak correlation between impact 
parameter and Mg II equivalent width (Churchill 1996) and the apparent 
absence of prominent absorption from gas-rich dwarf galaxies.
Although viewing galaxies via their absorbing effects provides a
valuable complement to the more traditional redshift surveys, it may
be that complex selection biases operate in such samples.  

On the other hand, it certainly is reassuring that the LFs of the
absorbers can be reconciled with the results emerging from the deep
redshift surveys. Impressive progress has been made in the past 2 years
from the comprehensive surveys of the CFRS group (Lilly et al 1995),
the LDSS/Autofib team (Colless et al 1990, Glazebrook et al 1995a, 
Ellis et al 1996b) and at the Keck (Cowie et al 1996). Collectively, the 
number of faint ($>$20 mag) redshifts is now over 1000 (Table 1) and each
provides a complementary insight into the distant population.

The CFRS survey is $I$-selected and well-suited for sampling the
evolving population of massive galaxies to $z\simeq$1. In contrast, the
LDSS/Autofib survey is $b_J$-selected and particularly tuned to address
the nature and distribution of the faint blue population which lies
around or fainter than $L^{\ast}$. The wide apparent magnitude range of
this survey makes it ideally suited for exploring changes in the {\it
shape} of the luminosity function with redshift. The unusually good
spectral resolution also makes it appropriate to examine evolutionary
trends as a function of spectral class (Heyl et al 1996). The
Keck survey by Cowie et al is $K$-selected and thus at high redshift is
least affected by uncertainties in $k$-corrections. Moreover, by using
multicolour data, Cowie et al have extended their survey in order to
construct $B$ and $I$ surveys to slightly deeper limits than has been
possible on 4-m telescopes.

\begin{table}[htb]
\caption[]{Deep Redshift Surveys}
\begin{center}
\renewcommand{\arraystretch}{1.2}
\begin{tabular}{llll}
\hline\noalign{\smallskip}
& Reference & $N_{gal}$ & Selection \\
\hline\noalign{\smallskip}
CFRS & Lilly et al (1995) & 591 & $I<$22 \\
LDSS & Ellis et al (1996b) & 1726 & 17$<b_J<$24 \\
Keck & Cowie et al (1996) & 346 & $K<$20 \\
& & 203 & $B<$24.5 \\                   
& & 130 & $I<$22.5 \\
\noalign{\smallskip}\hline\end{tabular}
\renewcommand{\arraystretch}{1}
\end{center}
\end{table}       

The empirical trends found by all 3 survey teams agree remarkably well
in the sense that the evolutionary changes seen are strongest in the
star-forming population which progressively occupy the fainter part of the
LF at lower redshift. Given the different perspectives and survey
strategies of the teams this is encouraging! Lilly et al (1996)
characterise the global evolution in terms of the mean rest-frame
luminosity density at various wavelengths and claim this corresponds
to an order of magnitude decrease in the volume-averaged star formation
rate since a redshift $z\simeq$1.

The difficulty lies in the physical interpretation of the declining
star formation rate in terms of the various populations and, in
particular, the question of whether number evolution is required. The
traditional `faint blue galaxy' problem has been sold as requiring an
`excess population' which fades or merges by $z\simeq$0 (e.g.  Ellis
1996). Is it possible to directly identify such an excess population
from the redshift surveys?

The CFRS team claim that the LF evolves such that the
most rapid change occurs for those galaxies with rest-frame colours bluer
than a typical Sbc. They discuss various galaxy populations whose 
characteristic evolutionary timescale differ. Number density evolution 
is not invoked. Although a very deep and well-controlled survey 
(the median redshift is $z\simeq$0.6), the {\it time} baseline is fairly
modest since there are few galaxies below a redshift of 0.3.

The LDSS team place greater emphasis on the changing {\it shape} of the
LF in the sense that the faint end slope steepens with increasing
redshift. Such a behaviour is not unexpected, at least qualitatively,
in hierarchical merging. By subdividing their large sample according to
[O II] strength and spectral class (Heyl et al 1996), they conclude
that the bulk of the evolution can be characterised by a strong
luminosity and/or number density evolution of the late type population.
Crucial to the need for number evolution is the assumed form for the
local LF.  A flat (Schechter $\alpha$=-1) local LF would imply fairly
dramatic changes have taken place between $z\simeq$0.5 and today, and
thus it is reasonable to question the reliability of the local LF 
before accepting this conclusion (McGaugh 1994).

Glazebrook et al (1995a), Ellis et al (1996b) and Cowie et al (1996)
have each placed constraints on the faint end slope of the local LF
from the absence of low $z$ galaxies in their deep $B$-selected surveys
and reject the hypothesis (Gronwall \& Koo 1995) that the counts can be
understood solely in terms of a local LF with a steep faint end slope.
On the other hand, it is clearly worrying that the same authors admit a
LF normalisation ($\phi^{\ast}$) markedly higher than traditional
estimates (e.g. Loveday et al 1992). Part of the difficulty may be the
degree to which brighter photographic photometry can be effectively
tied to that of the faint surveys (Bertin \& Dennefeld 1996) and this
is closely tied to the important role that surface brightness plays
in isophotal surveys (Ferguson \& McGaugh 1994). 

What evidence is there for short-term star formation activity such as
might be expected if merging is an ingredient driving the evolution?
Broadhurst et al (1988) first suggested that the faint blue
star-forming sources had spectral characteristics indicating short-term
bursts rather than a gradual decline in the star formation rate. More
recently, Heyl et al (1996) have found a similar effect in the more
extensive LDSS and Autofib survey (Figure 2). In the blue-selected
samples, a high proportion of the spectra are unlike those of local
spirals. For this class of object there is also a marked increase in
the median [O II] 3727 \AA\ equivalent width with redshift (Figure 3).

\begin{figure}
\vspace{5.5cm}
\caption{Coadded spectra for late-type spiral galaxies
in the LDSS/Autofib survey (Heyl et al 1996). The bold curve is
the coaddition of those with $z<$0.2 while the light curve is for
those with 0.2$<z<$0.5. The higher redshift sample shows absorption 
lines whose increased strength is indicative of recent 
($\simeq$1 Gyr) star formation.} 
\end{figure}

\begin{figure}
\vspace{4.5cm}
\caption{Evolution of the median [O II] equivalent width
for late-type spirals in the LDSS/Autofib survey (Heyl et al 1996).} 
\end{figure}

Assuming galaxy morphology is a marker with some degree
of permanence, HST galaxy counts can provide valuable insight into
the galaxies that are responsible for the LF changes discussed above.
Glazebrook et al (1995b) and Abraham et al (1996b) provide type-dependent
counts to $I$=25 from the Medium Deep Survey and the Hubble Deep Field
and claim a remarkable overabundance of irregular galaxies compared
to local samples. Many are certainly suggestive of merging, although 
firm quantitative proof is difficult to establish (Neuschaefer
et al 1995). Although these counts probe much deeper than the current
redshift survey limits, the basic conclusion is that the number of
regular spheroidal and disk galaxies to $I\simeq$22-23 is fairly close
to no evolution expectations, whereas the bulk of the excess population
seems confined to the irregular sources.

But how reliable are the morphological assignments in these faint
samples? Abraham et al (1996c) have addressed this question via the
development of more quantitative classification criteria based on
assymmetry and light concentration and via simulations that take
account, on a pixel-by-pixel basis, of differential $k$-corrections and
surface brightness dimming. For z$<$1, where the bulk of the $I<$22
galaxies lie, the biases are quite small and amount, for example, to
confusion only between Sdm, Irregular or merging systems rather than
gross misclassification such as movement from Sbc to Irr.  As local
irregulars and late type spirals should not be difficult to see if they
are still actively producing stars, their abundance in the HST counts
compared to a paucity in local data is an important result.

The HST and ground-based data thus both assign a high proportion of 
the evolution to a single class, namely late-type spirals and irregulars.
The mean luminosity and perhaps number density of this class of sources
is rapidly decreasing with time. This is not to say there is not room
for some evolution in the intermediate spirals or ellipticals.
The CFRS and LDSS teams have joined forces and will soon have over 300 
galaxies for which HST images and spectra will be available to $I$=22 
and $b_J$=24. Schade et al (1996) have already claimed quite 
significant evolution in the surface brightness of disk galaxies 
with redshift and this work is being extended to the larger CFRS+LDSS 
database. Additionally, the luminosity and redshift distribution of morphologically-distinct samples is being analysed in the context
of the question of whether number evolution is required.

The most significant conclusion thus far from the redshift
surveys is the global evolution of the population (Lilly et al 1996)
rather than results based upon dissection into individual types whose 
physical significance remain unclear. Nonetheless, for a detailed 
understanding of the origin of disk galaxies and a resolution of the 
`faint blue galaxy problem' it is clear this is the way to go although
possibly very large joint HST and ground-based spectroscopic samples
will need to be gathered.

\section{Exploring $z>$1}

The Hubble Deep Field has provided an exciting first glimpse into the
distant Universe. By extending the galaxy counts to a regime
affected by ground-based confusion, it is now clear that the surface 
density of very faint sources exceeds that predicted on the basis of
most reasonable local luminosity functions, suggesting either:
(i) galaxies are not conserved, viz. in the HDF we are seeing
sub-units destined to become larger systems, or (ii) we live in
a world model where $\Lambda\neq$0.

Although $\Lambda$-dominated models remain popular in theoretical
circles, the observational constraints are getting tighter.  If the
excess counts were produced primarily by huge volumes, one would see
similar effects in the $K$-band (Djorgovski et al 1995). Spatially-flat
models with $\Lambda\neq$0 would also produce accelerating universe
which seem in conflict with the constraints emerging from the
distant supernovae programmes (Perlmutter et al 1996, Leibundgut, this
volume).

There is growing evidence that the bulk of the star formation that made
the present day population occurred between 1$<z<$3. Firstly, the steep
rise in star formation rate to redshift $z\simeq$1 is a major pointer
to a low redshift of mean star formation (Fall et al 1996). Secondly, a
key result, suggested initially by Guhathakurta et al (1990), is the
rarity of $R<$25 star-forming sources whose $U$ band flux indicates a
strong Lyman limit consistent with $z>$3. The same result has been
derived by Steidel and collaborators with the important advance that
candidate Lyman limit sources beyond $z\simeq$2.3-3 have now been
spectroscopically verified using the Keck telescope both in the HDF
(Giavalisco et al 1996) and elsewhere (Steidel et al 1996).
 
If the bulk of the star formation occurred at low redshift, could the
high $z$ star-forming galaxies recently identified be the ancestors of
the present day spheroidal galaxies? HST images of those HDF galaxies
satisfying Abraham et al's (1996c) Lyman-limit criteria are shown in
Figure 4 and are highly suggestive of hierarchical merging of sub-units
in the manner predicted by Kauffmann (1996) and Baugh et al (1996).
However, a crucial pointer in this regard would be an estimate of
the mass of such distant systems, either from resolved spectroscopy
or absorption line widths (c.f. Giavalisco, this volume).

\begin{figure}
\vspace{13.5cm}
\caption{Hubble Deep Field images of those galaxies
selected by Abraham et al (1996b) to have Lyman limit drop outs
suggesting they lie at redshifts $z>$2.3. A significant fraction
have since been spectroscopically confirmed (Giavalisco et al 1996).} 
\end{figure}

It is important to recognise that the Lyman-limit method, although
remarkably effective, provides only a limited view of the high
redshift Universe, namely star-forming sources within a narrow redshift
range. Our inability to immediately `connect' this interesting
population with low $z$ counterparts exemplifies the need to provide
complete redshift coverage so the evolution of the various samples can
be directly tracked.

How are we going to {\it systematically} explore the galaxy population
beyond $z\simeq$1 in a manner similar to that which has been so
successful with the 4-m telescopes for $z<$1?  If our hypothesis
concerning the star formation history is correct, the redshift interval
1$<z<$2.5 is particularly important yet, paradoxically, this is a range
which will be the hardest to systematically explore with the first
tranche of 8-m instruments.  The basic difficulty is the absence, at
optical wavelengths, of any of the familiar diagnostic features.
Although one is encouraged by the detection of absorption lines in the
Keck spectra to faint limits, I am confident those same sources would be far
easier to study at infrared wavelengths where their emission line
spectra would be quite prominent.

At this conference we have witnessed three promising techniques.
Firstly, the weak lensing signals seen in a variety of clusters provide
valuable information on the mean statistical distance to an objective
sample of very faint images viewed through the cluster lens (Fort, this
volume). In certain clusters, the mass models are sufficiently tightly
constrained (Kneib et al 1996) that the modelling can be used to estimate
distances to very faint sources viewed through the lens. The process
can be iteratively improved via arclet redshift measurements to make 
quite precise statements about extremely faint galaxies. A dramatic 
verification of this `inversion' technique is the $z$=2.515 arc in Abell 2218 
(Ebbels et al 1996, Figure 5) originally predicted to be a
$R_{true}$=24.1 galaxy at $z\simeq 2.8\pm0.3$. This is just the beginning 
of several approaches which, through calibration spectroscopy and HST 
imagery, will provide mean distances to a variety of galaxy populations at 
very faint limits. As these methods are geometric, they avoid many of the 
biases inherent in traditional surveys.

\begin{figure}
\vspace{6.0cm}
\caption{Spectrum of a faint arc (\#384) in the rich cluster Abell 2218
obtained with LDSS-2 on the WHT (Ebbels et al 1996). The spectroscopic
redshift of 2.515 agrees closely with that inferred from the lensing
inversion method developed by Kneib et al (1996) and illustrates the
potential of determining the mean redshift of a very faint population
of galaxies viewed through a lensing cluster.} 
\end{figure}

Secondly, a high priority must be the effective use of the panoramic
infrared spectrographs on 8-10m telescopes to sample the wavelength range where
redshifted [O II] and $H\alpha$ lie. The troublesome OH background
necessitates high dispersion which is costly in detector pixels.  The
VIRMOS project (LeFevre, this volume) is an imaginative and ambitious
solution to this problem. The third technique is the highly
complementary CADIS narrow band imaging programme (Meisenheimer, this
volume). On the relative merits of these two techniques, I believe it
will be important to execute some scouting missions to determine the
likely distribution of emission line strengths before finalising the
design parameters of a major commitment like VIRMOS. One would hardly
contemplate embarking on 2dF or the Sloan Digital Sky Survey without
having gathered a representative set of optical galaxy spectra.

\section{Acknowledgements} I thank Jacqueline Bergeron and Christina
Stoffer for their generous financial assistance which made it possible
for me to attend this enjoyable meeting. I acknowledge useful
conversations with Carlos Frenk, Simon Lilly, Guinevere Kauffmann and
Alvio Renzini. I thank all my collaborators on the lensing and cluster
and field galaxy programmes for allowing me to discuss our work in the
context of this review.

\smallskip
 
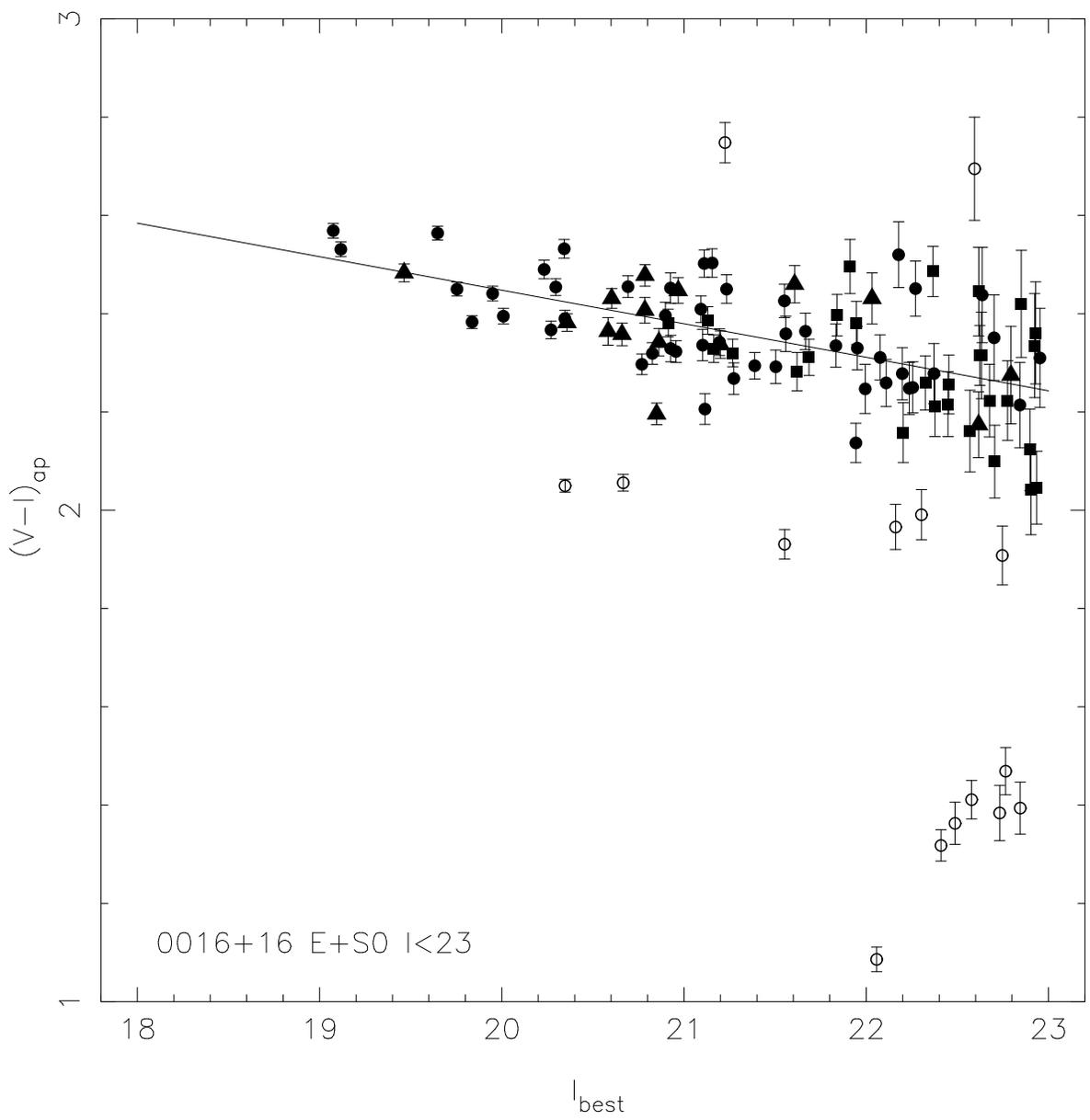
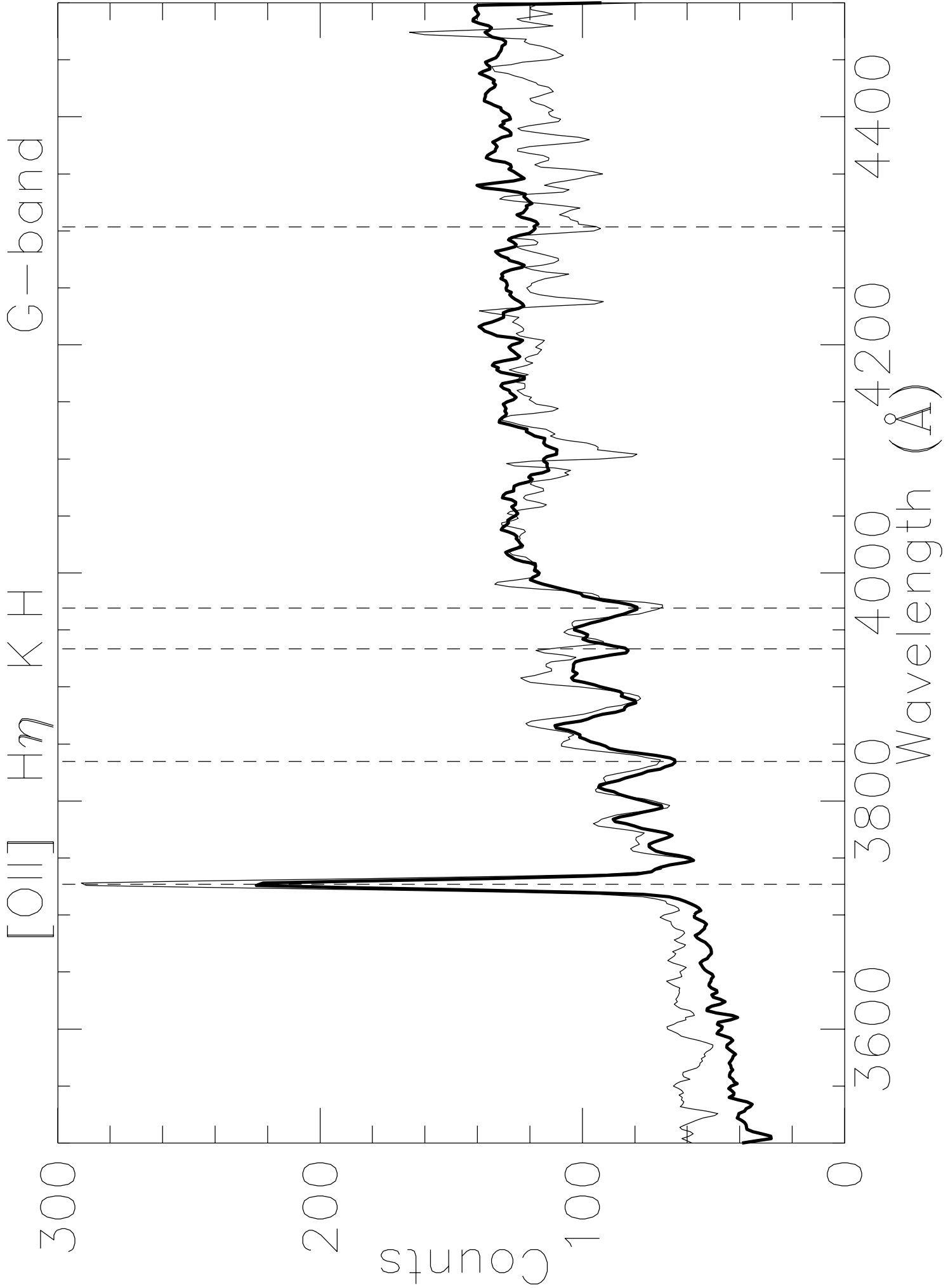
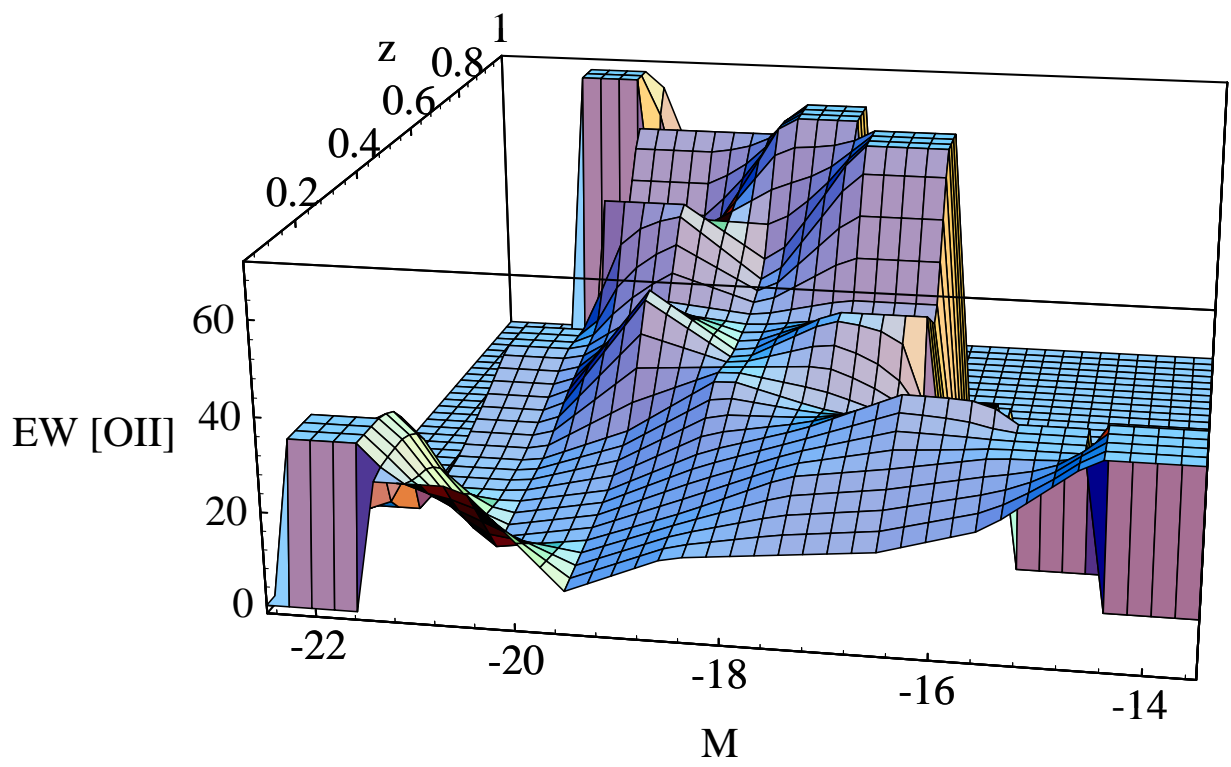
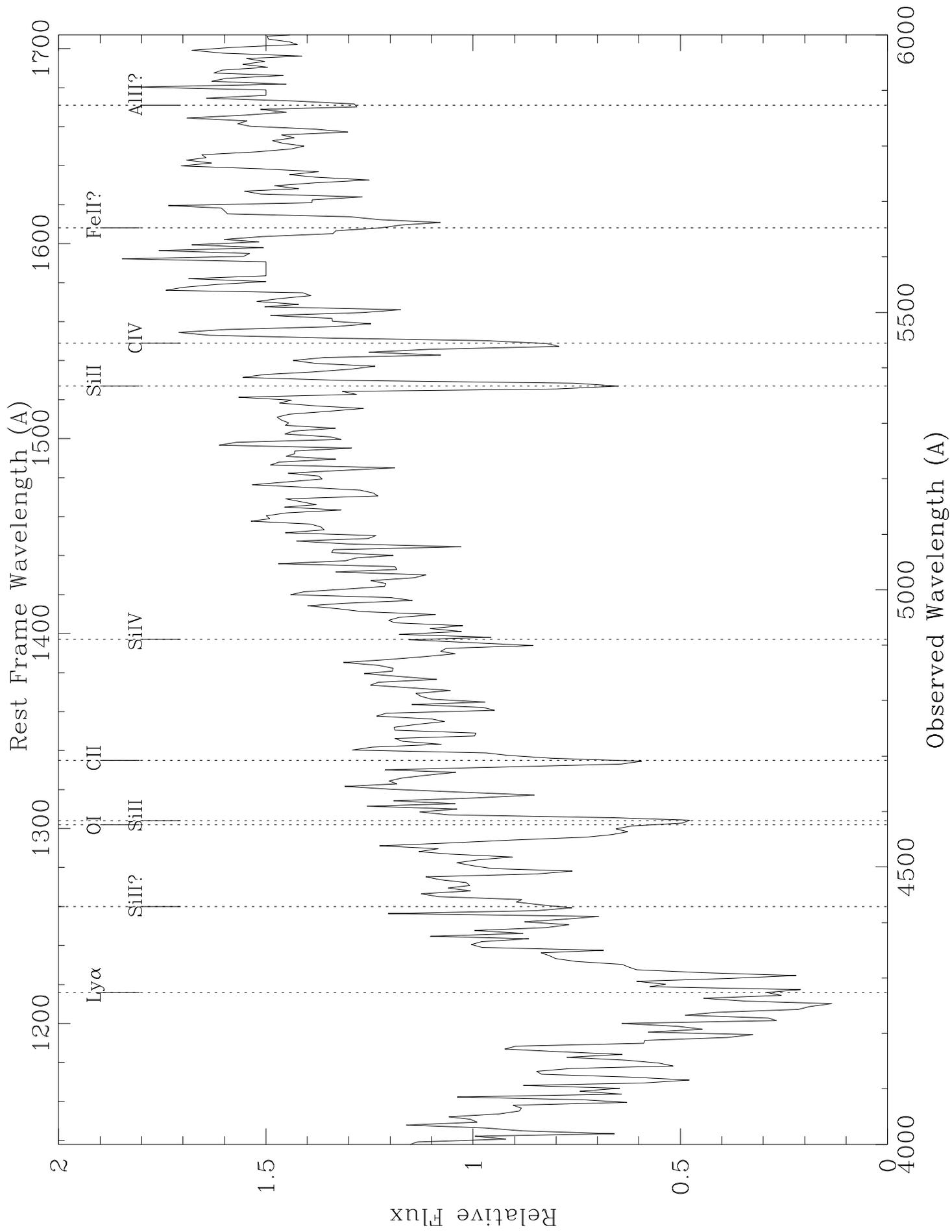
\end{document}